\documentstyle[12pt]{article} 
\font\elevenbf=cmbx10 scaled\magstep 1                                        
\begin{document}                                                              
%\begin{flushright}{LNF-94/xxx P}  
%\end{flushright}                                                      
\begin{center}                                                                
{ \large\bf Quantum Instability for Mixed States \\}
\vskip 2cm  
{ Marek Nowakowski \\}                              
Instituto de Fisica, Universidad de Guanajuato, Apdo Postal E-143,
Leon, Guanajuato, Mexico
\end{center}
\vskip .5cm
\begin{center}
\end{center}                             
\begin{abstract}
The analysis of the time evolution of unstable states which are
linear superposition of other, observable, states can, in principle, be 
carried out in two distinct, non-equivalent ways. 
One of the methods, usually employed for the neutral kaon system, 
combines the mixing and instability into
one single step which then results into unconventional properties of the
mass-eigenstates. An alternative method is to remain within the framework of
a Lagrangian formalism and to perform the mixing prior to the instability 
analysis. Staying close to the  $K^0-\bar{K^0}$ system, we compare both 
methods pointing out some of their shortcomings and advantages.  
\end{abstract}                                                                
\newpage 
\section{Introduction}
Consider a single unstable states $\vert \lambda \rangle$ characterized by some
quantum numbers denoted by $\lambda$. 
Although not an eigenstate to the full Hamiltonian, the state has a definite
mass.
Viewing this state as an open quantum
system, its time evolution can be generally given as
\begin{equation} \label{inst}
\vert \lambda (\tau) \rangle 
= p_{\lambda}(\tau)\vert \lambda \rangle
\end{equation}
where the effective Hamiltonian ${\cal H}_{\rm eff}$ governing this system
is non-hermitian and in the Weisskopf-Wigner approximation 
\cite{ww} one recovers the
exponential decay law $p_{\lambda}(\tau)=e^{-iE\tau}e^{-\Gamma/2 \tau}$.
As long as we restrict ourselves to a single particle state $\vert \lambda \rangle$,
there are no known inconsistencies in performing this analysis. The
state $\vert \lambda \rangle $ can be a mass-eigenstate defined, say, through
mixing in a Lagrangian framework. Suppose now that we combine this mixing
and the instability analysis into one single equation. With a two level 
quantum system, the non-hermitian Hamiltonian of the single particle case
becomes a non-hermitian effective $2 \times 2$ mass matrix 
${\cal M}_{\rm eff}$. Obviously, the diagonalization of 
${\cal M}_{\rm eff}$ is
generally not given by a unitary transformation and as a result the norm is 
not automatically preserved. The two emerging mass eigenstates, say $\vert
\lambda_{1,2} \rangle$, can turn out to be non-orthogonal i.e. $\langle
\lambda_1 \vert \lambda_2 \rangle \neq 0$. This non-orthogonality has rather
profound consequences: no suitable definition of the anti-particle states 
is known, the time development of $\vert \lambda_1 \rangle$ and
$\vert \lambda_2 \rangle$ are generally inter-connected (that is to say
that the times evolution of one of this states seems to `know` 
about the other) and, maybe less certain, EPR-like paradoxes can be 
encountered in the system. A more detailed account of these peculiarities will
be given in section 3. 
Let us now assume that the mixing is well defined within a Lagrangian. 
Performing only unitary transformations we can define two 
orthogonal mass eigenstates
and proceed to 
analyze their time behaviour for each one of them separately.
Since the situation of mixing and instability is often 
encountered in physics, the rather fundamental question arises, which one of 
the possibilities we should follow. Obviously, the outcomes will be different.
The first method is followed in the neutral kaon system 
\cite{loy, kabir} (and 
related mesonic systems) whereas for 
massive unstable neutrinos the mixing is done within the Langrangian framework \cite{petcov} and
we opt here for the second possibility. Now the formal difference is 
that for the
weak, CP-violating, kaon 
interaction no effective Langrangian description is known
(this does not mean that such description is in general 
not possible and indeed we present below a Lagrangian which, at least,
theoretically, contains the basic features of the neutral kaon system) whereas
the neutrinos are part of the fundamental extended Standard Model Langrangian.
On the other hand, there exist no argument which would restrict the 
applicability of the first method only to kaons or more generally to composite
objects. 
Indeed, regardless whether
the particles are fundamental or composite, this method can be applied  to any
unstable two level system, in particular to neutrinos. Since the question we 
pose here touches upon a basic principle, 
we should be able to decide which one is physically correct.
If we find that the first method is the preferable one, then it should be 
applied to any similar system even if this system has been initially
formulated by means of a Lagrangian. On the other hand if we find this method
erroneous, we should seek for better theoretical tools which could amount
to introduce improvements or to abandon it and replace it e.g. by a suitable
effective Lagrangian. 

In this paper we will compare these two approaches without propounding 
one in favor of the other. In doing so, we will stay as close as possible
to the  $K^0-\bar{K^0}$ system. Although the Lagrangian presented below
captures many features of the neutral kaon system, this will not be the main
issue of the paper. As indicated above, no condition on the elementarity or
compositness of the involved states enters the 
Lee-Oehme-Yang (LOY) theory \cite{loy} which is
the theory which combines mixing of  $K^0-\bar{K^0}$ with the instability of 
these states. That is to say that this theory goes formally 
through for composite as 
well as fundamental particles. Therefore, to emphasize the main conclusions
of the paper we will mostly treat the states as elementary. Nevertheless
a comparison to the  $K^0-\bar{K^0}$ system seems legitimate.

\setcounter{equation}{0}
\section{CP-violation through mixing in a Lagrangian}
To facilitate the understanding of the main problem, we start with
a simple Lagrangian involving two complex scalar fields $\phi$ and $\chi$, one 
scalar neutral field $\varphi$ and possibly other fields $\eta_i$ which
we do not need to specify further. We have
\begin{equation} \label{lagrangian}
{\cal L}=
{\cal L}^{\rm kin}_{\varphi}+
{\cal L}^{\rm kin}_{\phi}+
{\cal L}^{\rm kin}_{\chi}+
{\cal L}^{\rm int}_{\phi \varphi \chi}+
{\cal L}^{\rm mix}_{\rm CP-violat.}+
{\cal L}^{\rm int}_{\phi \varphi \chi \eta_i}
\end{equation}
where ${\cal L}^{\rm kin}$ are the usual kinetic terms of the same form as
\begin{equation} \label{kin}
{\cal L}^{\rm kin}_{\phi}=(\partial_{\mu}\phi)(\partial^{\mu}\phi^*) -
m^2\phi \phi^*
\end{equation}
We choose for the interaction between $\phi$, $\varphi$ and $\chi$ the simple
expression
\begin{eqnarray} \label{int}
{\cal L}^{\rm int}_{\phi \varphi \chi} &=& {\lambda_{00} \over \sqrt{2}}
\phi \varphi^2 + {\lambda_{+-} \over \sqrt{2}}\phi \chi \chi^* \nonumber \\
&-& i{\lambda_{000} \over \sqrt{2}} \phi \varphi^3 -i{\lambda_{+-0}
\over \sqrt{2}}\chi \chi^* \varphi + {\rm h.c.}
\end{eqnarray}
where $\lambda_{00}$, $\lambda_{+-}$, $\lambda_{000}$ and $\lambda_{+-0}$ are
real coupling constants. We demand now that 
${\cal L}^{\rm int}_{\phi \varphi \chi \eta_i}$ be by itself CP-invariant
such that $\varphi$ has the CP quantum numbers of a pseudoscalar.
Furthermore, we impose on ${\cal L}^{\rm int}_{\phi \varphi \chi \eta_i}$
a global $U(1)$ symmetry 
such that among $\phi$, $\varphi$ and $\chi$ only $\phi$ has a non-zero charge
with respect to this global $U(1)$ symmetry. 
Then ${\cal L}^{\rm int}_{\phi \varphi \chi }$
breaks this symmetry and even with 
${\cal L}^{\rm mix}_{\rm CP-violat.}$ being zero we cannot identify the
$\phi$ and $\phi^*$ as mass-eigenstates. Writing therefore 
$\phi=(\phi_1 +i\phi_2)/\sqrt{2}$ where $\phi_{1,2}$ are now the proper mass
eigenstates (no mixing yet) we obtain easily
\begin{eqnarray} \label{int2}
{\cal L}^{\rm int}_{\phi \varphi \chi} &=& \lambda_{00} 
\phi_1 \varphi^2 + \lambda_{+-} \phi_1 \chi \chi^* \nonumber \\
&+& \lambda_{000} \phi_2 \varphi^3 +\lambda_{+-0}
\phi_2 \chi \chi^* \varphi
\end{eqnarray}
The Lagrangian 
\begin{equation} \label{lagrangian1}
{\cal L}^{(1)}= {\cal L} -{\cal L}^{\rm mix}_{\rm CP-violat.}
\end{equation}
in which the mass-eigenstates $\phi_1$ and $\phi_2$ 
are defined, describes the interaction
of four mass-eigenstates fields: the neutral pseudoscalar (CP-odd) 
fields $\varphi$, the
charged field $\chi$ and the scalar (CP-even) 
$\phi_1$ as well as the pseudoscalar
$\phi_2$. Obviously, since we were able to assign CP quantum numbers to all
fields ${\cal L}^{(1)}$ is still invariant under CP-transformations. How can
we break the CP-invariance through mixing of $\phi_1$ and $\phi_2$ \cite{me1}? 
To this end 
we introduce 
\begin{eqnarray} \label{mix}
{\cal L}^{\rm mix}_{\rm CP-violat.} &=& -\mu^2 \phi \phi - \mu^{*2}\phi^*
\phi^* \nonumber \\
&=& -\sqrt{2} \Re {\rm e} \mu^2 \phi_1^2 + \sqrt{2} \Re {\rm e} \mu^2 \phi_2^2 -
\sqrt{2} \Im {\rm m} \mu^2 \phi_1 \phi_2
\end{eqnarray}
with a complex parameter $\mu^2$. With $\Im {\rm m} \mu^2 \neq 0$, the fields
$\phi_1$ and $\phi_2$, previously carrying the quantum numbers of scalar and
pseudoscalar, respectively, will mix which evidently leads to CP-violation.
With $\Im {\rm m} \mu^2 =0$, there is no CP-violation, but the mixing remains. 
Hence, $\phi_1$ and $\phi_2$ are no longer mass-eigenstates. 
As a side remark, we note that ${\cal L}^{\rm mix}_{CP-violat.}$ can be 
understood as describing the transition $\phi \leftrightarrow \phi^*$.
With a little bit effort, one could obtain at higher orders 
such a transition from
${\cal L}^{(1)}$ in (\ref{lagrangian1}) since we have $\phi \leftrightarrow
\chi \chi^* \leftrightarrow \phi^*$. However, because ${\cal L}^{(1)}$
respects the CP-symmetry (at least with all coupling constants real),
the contributions would be also CP-invariant.

The mixing due to (\ref{mix}) is given by
\begin{equation} \label{mixing}
-{1 \over 2}
\left(\begin{array}{c}
\phi_1 \\
\phi_2
\end{array}\right)^T
\left(\begin{array}{cc}
m^2+2 \Re {\rm e} \mu^2 & -2 \Im {\rm m} \mu^2 \\
-2\Im {\rm m}  \mu^2 & m^2 - 2\Re {\rm e} \mu^2
\end{array}\right)
\left(\begin{array}{c}
\phi_1 \\
\phi_2
\end{array}\right)
\end{equation}  
which after diagonalization leads to two new mass-eigenstates, $\Lambda_1$
and $\Lambda_2$
\begin{eqnarray} \label{massstates}
\Lambda_1 &=& {1 \over \sqrt{2}}e^{i\theta}[\phi + e^{-2i\theta}\phi^*]
\nonumber \\
&=& \cos \theta \phi_1 - \sin \theta \phi_2
\nonumber \\
\Lambda_2 &=& {1 \over \sqrt{2}}e^{i\theta}[\phi - e^{-2i\theta}\phi^*]
\nonumber \\
&=& \sin \theta \phi_1 + \cos \theta \phi_2
\end{eqnarray}
with masses given by
\begin{eqnarray} \label{mass}
m_{\Lambda_1}^2 &=& m^2 + 2 \vert \mu^2 \vert 
\nonumber \\
m_{\Lambda_2}^2 &=& m^2 - 2 \vert \mu^2 \vert
\end{eqnarray}
The mixing angle $\theta$ is defined through
\begin{equation} \label{theta}
\tan 2 \theta ={\Im {\rm m} \mu^2 \over \Re {\rm e} \mu^2}
\end{equation}
In terms of the new mass-eigenstates $\Lambda_{1,2}$ the interaction
Lagrangian ${\cal L}^{\rm int}_{\phi \varphi \chi}$ reads 
\begin{eqnarray} \label{int3}
{\cal L}^{\rm int}_{\phi \varphi \chi} 
&=& \lambda_{00}
(\cos \theta \Lambda_1 + \sin \theta \Lambda_2 )\varphi^2
\nonumber \\
&+& \lambda_{+-}
(\cos \theta \Lambda_1 + \sin \theta \Lambda_2 )\chi \chi^*
\nonumber \\
&+& \lambda_{000}
(-\sin \theta \Lambda_1 + \cos \theta \Lambda_2 )\varphi^3
\nonumber \\
&+& \lambda_{+-0}
(-\sin \theta \Lambda_1 + \cos \theta \Lambda_2 )\chi \chi^* \varphi
\end{eqnarray}
Performing once again a CP-transformation, but now on mass-eigenstates,
we see that the CP-symmetry is broken as long as $\sin \theta \neq 0$
i.e. $\Im {\rm m} \mu^2 \neq 0$. With CP-violation and provided that 
$m_{\Lambda_{1,2}}$ are bigger than $2m_{\varphi}$ and $m_{\varphi}
+ 2m_{\chi}$, both states 
$\Lambda_1$ and $\Lambda_2$ will simultaneously decay
into two, i.e. $\varphi \varphi$ and $\chi \chi^*$, and three, i.e.
$\varphi \varphi \varphi$ and $\chi \chi^* \varphi$, spinless states.

There are two aspects of the Lagrangian presented in equations
(\ref{lagrangian})-(\ref{int3}). We postpone the discussion of the more
important aspect to the section 3 after having briefly discussed in section 4
the Lee-Oehme-Yang theory. 

The Lagrangian (\ref{lagrangian}) captures, at least theoretically, all 
important features of the $K^0-\bar{K^0}$ system and is applicable to 
fundamental as well as composite fields (in the latter case we should
replace the coupling constants by form factors, but the essential
conclusions would remain unchanged). Indeed, we can identify 
$\phi$ and $\phi^*$ with
$K^0$ and $\bar{K^0}$, respectively and the mass $m$ with the kaon mass
parameter usually denoted by $m_K$. The field $\varphi$ represents then the
neutral pion and $\chi$ the charged pion field. The other fields
$\eta_i$ in 
${\cal L}^{\rm int}_{\phi \varphi \chi \eta_i}$ stand, of course
for more hadronic states. Hence 
${\cal L}^{\rm int}_{\phi \varphi \chi \eta_i}$ is the strong interaction part
which conserves CP and strangness, our imposed $U(1)$ symmetry. The latter
is broken by weak interaction i.e. by 
${\cal L}^{\rm int}_{\phi \varphi \chi}$. If $\Im {\rm m} \mu^2 =0$ i.e. there is no
CP-violation in the system, $\Lambda_1$ and $\Lambda_2$ have the quantum
numbers of scalar and pseudoscalar and as such can be identified with
$K_1$ and $K_2$ which do not decay simultaneously into two and three pions.
If $\Im {\rm m} \mu^2 \neq 0$, the CP-violation makes, however, both decays possible.
A direct CP-violation can be also switched on by making the coupling
constants in 
${\cal L}^{\rm int}_{\phi \varphi \chi}$ complex. Anticipating our later 
discussion, we point out two essential features of 
the Lagrangian (\ref{lagrangian}). Unlike the $K^0-\bar{K^0}$ system
described in the LOY-theory, the mass-eigenstates in the Lagrangian
remain orthogonal and this in spite of CP-violation. Secondly, $\Lambda_1$
($\Lambda_2$) has a well-defined anti-particle, namely it is anti-particle
to itself.
 
The above is only a comparison. It would be too early to take 
(\ref{lagrangian})
as a viable phenomenological description of the $K^0-\bar{K^0}$ system. One 
reason is certainly that since this proposal is new, we do not know how to make
a connection between the phenomenological Lagrangian (\ref{lagrangian})
and the more fundamental Lagrangian of the Standard Model at quark level.
This does not mean that such a connection cannot be established in future.
If theoretically such a connection turns out to be possible, it could still be
that (\ref{lagrangian}) cannot reproduce all experimental facts from
the $K^0-\bar{K^0}$ system in which case we are left only with the choice
of the LOY-theory. 
Secondly, to investigate the semi-leptonic decay
channels with (\ref{lagrangian}), 
we would have to add to (\ref{lagrangian}) a $W-\Lambda_{1,2}-
\varphi (\chi)$ interaction which as such is actually not a problem. However, 
at tree level we would not see any CP-violation in the semi-leptonic
decays and to obtain a non-zero semi-leptonic CP-asymmetry higher order
corrections would be necessary. Since this superficial identification of
(\ref{lagrangian}) with the neutral kaon system is not the main point
of the paper, we will not dwell on it further here. We end this section
by once again pointing out that the analysis of the time evolution
of $\Lambda_{1,2}$ can be now done for each individual mass-eigenstate
separately.
\setcounter{equation}{0}
\section { The LOY theory}
To compare the results from the previous section, we will outline now
the basic points of the LOY theory \cite{loy, kabir}. 
Although we will formulate it for the 
neutral kaon system, it is important to stress again that this theory
is applicable to many unstable systems with mixing, be it elementary or
composite. In particular, it would also apply to massive neutrinos and the
theory whose Lagrangian we presented in section 2. The LOY theory examines
the mixing and instability simultaneously and as a result of this analysis
we end up with an effective non-hermitian mass matrix ${\cal M}_{\rm eff}$
which replaces the effective Hamiltonian for a single unstable state.
This mass matrix is given by
\begin{eqnarray} \label{massmatrix}
({\cal M}_{\rm eff})_{ij} &=& m_K \delta_{ij} + \langle K^0_i
\vert H_{\rm weak} \vert K^0_j \rangle \nonumber \\
&+& \sum_n {\langle K^0_i \vert H_{\rm weak}\vert n \rangle
\langle n \vert H_{\rm weak} \vert K^0_j \rangle \over 
m_K -E_n +i\epsilon}
\end{eqnarray}
where $K^0_i$ can be either $K^0$ or $\bar{K^0}$.
Parametrizing this mass matrix by
\begin{equation} \label{para}
{\cal M}_{\rm eff}=
\left(\begin{array}{cc}
A & p^2 \\
q^2 & A
\end{array}\right)
\end{equation}
the mass-eigenstates can be calculated to be
\begin{eqnarray} \label{KSL}
\vert K_{S/L}\rangle &=& p\vert K^0 \rangle  \pm q \vert \bar{K^0} \rangle
\nonumber \\
\vert p\vert^2 &+& \vert q \vert^2 =1
\end{eqnarray}
Because ${\cal M}_{\rm eff}$ is in general non-hermitian, it is not
necessary that we have $\langle K_S \vert K_L \rangle = \langle K_S \vert
K_L \rangle =0$ or, in other words, that $\vert p \vert^2 =\vert q \vert^2$.
Indeed, if CP is a good symmetry, it follows that
\begin{equation} \label{CPkaon}
p^2 = \langle K^0 \vert {\cal M}_{\rm eff} \vert \bar{K^0} \rangle
=e^{-2i\xi}\langle \bar{K^0} \vert {\cal M}_{\rm eff}\vert
K^0 \rangle = e^{-2i\xi}q^2
\end{equation}
where the phase $\xi$ comes from the CP-transformation $CP \vert
K^0 \rangle = e^{i\xi}\vert \bar{K^0} \rangle$. We conclude that
if $\vert p \vert^2 -\vert q\vert^2 \neq 0$, we have CP-violation in 
the LOY theory and, indeed, this is taken as a genuine signal of
broken CP. However, if $p^2=e^{i\beta}q^2$, with some phase $\beta$,
we cannot decide whether CP is broken or conserved. As apparent
from the Lagrangian example in section 2, a system can still
exhibit CP-violation even if $\vert p \vert^2 =\vert q \vert^2$.
Quite similarly, if $K_{S/L}$ are eigenstates to CP, then 
$\vert p \vert^2 =\vert q \vert^2$ (or if 
$\vert p \vert^2 =\vert q \vert^2$, 
then $K_{S/L}$ are eigenstates to CP). The assignment 
or non-assignment of CP quantum numbers to the mass-eigenstates 
within the LOY-theory occurs in separation from any other state. In contrast,
in the Lagrangian case this happens in conjunction with all states involved
(i.e. only if we can assign to all fields a CP-phase, is CP conserved). As
said above with $\vert p \vert^2 =\vert q \vert^2$ CP can still be violated 
in the Lagrangian approach.

The time development in the Weisskopf-Wigner approximation 
is given by the the Schr\"odinger-like equation
\begin{equation} \label{time}
i{d \over d\tau}
\left(\begin{array}{c}
\vert K^0 (\tau)\rangle \\
\vert \bar{K^0}(\tau)\rangle
\end{array}\right)
= {\cal M}_{\rm eff}
\left(\begin{array}{c}
\vert K^0 (\tau)\rangle \\
\vert \bar{K^0}(\tau)\rangle
\end{array}\right)
\end{equation}
which leads to the exponential decay law for the mass-eigenstates.

In conclusion, in the LOY theory analyzing mixing and instability in a single
step, one has to handle a non-hermitian mass-matrix. 
Its effect, in the presence
of CP-violation, is the non-orthogonality of the mass-eigenstates i.e.
\begin{equation} \label{nonortho}
\langle K_L \vert K_S \rangle = \langle K_S \vert K_L \rangle
= \vert p\vert^2 -\vert q \vert^2 \neq 0
\end{equation}
In the next section we will discuss some peculiarities connected with
non-orthogonal states. As such (\ref{nonortho}) is already troublesome
to interpret. Although widely used, it is at the same time sometimes
doubted if the usual projection technique in quantum mechanics is still
applicable here. A density matrix would be preferable, but as for now
one treats the mass-eigenstates kaons as pure states i.e. the density matrix
is $\rho_{S/L}=\vert K_{S/L}\rangle \langle K_{S/L} \vert$ and then there is 
no essential difference to the projection technique.

It should not come now as a surprise that in LOY theory CP-violation
is correlated with the width of $K_{S/L}$. The best place to see it, is
by means of the Bell-Steinberger relation which reads \cite{bell}
\begin{eqnarray} \label{BS}
(\lambda_L -\lambda_S^*) \langle K_S \vert K_L \rangle &=&
\sum_f \langle f\vert T \vert K_S \rangle^*\langle f \vert T \vert
K_L \rangle 
\nonumber \\
\lambda_{S/L} &\equiv &m_{S/L} -{i \over 2}\Gamma_{S/L}
\end{eqnarray}
where $T$ is the transition operator. Evidently, if the transition matrix 
elements containing $K_S$ or $K_L$ vanish (which is equivalent to 
having $\Gamma_S$ or $\Gamma_L$ zero) and maintaining $m_S \neq m_L$, we are 
forced to assume that $\langle K_S \vert K_L \rangle =0$ i.e. no CP-violation.
Again this is not so in the Lagrangian example where even if the 
$\Lambda_{1,2}$ mass-eigenstates are stable, the CP-violation does not
vanish and, as a matter of principle, could show up in a different place
(e.g. in scattering processes). There is no such correlation between the 
mixing parameters and the width in the Lagrangian formalism.

An important point to realize is that the LOY theory is not restricted to the
kaon system. Provided we are ready to leave the 
Lagrangian formalism, it would go through also for the interaction presented
in section 2. Indeed, we would not even need to know the explicit form
of 
${\cal L}^{\rm int}_{\phi \varphi \chi}+ {\cal L}^{\rm mix}_{\rm CP-violat.}$
corresponding now to $H_{\rm weak}$. All that is 
formally required, is  the definition
of $\phi$ and $\phi^*$ through ${\cal L}^{\rm int}_{\phi \varphi \chi \eta_i}$
and the assumption that  
${\cal L}^{\rm mix}_{\rm CP-violat.}$ violates CP. 
We could also demand that the interaction contained in 
${\cal L}^{\rm int}_{\phi \varphi \chi \eta_i}$ is much stronger than in
the corresponding ${\cal L}^{\rm int}_{\phi \varphi \chi}$.
As compared with the
results from the Lagrangian, this would lead to non-orthogonal 
mass-eigenstates. We emphasize that this orthogonality is not due to some
artifact of an approximation, but essentially a consequence of the 
non-hermiticity of the effective mass matrix. This property of
${\cal M}_{\rm eff}$ will be always present in the analysis of the time
evolution of a two level system.

Certainly, the advantage of the LOY theory is that there exit a well 
established
connection to the Standard Model Lagrangian and, even more importantly, 
the experimental data. Nevertheless, it has some curious theoretical 
properties which we shall examine in the next section.

\setcounter{equation}{0}
\section{Consequences of non-orthogonality}
We have seen that for the dynamics defined through the Lagrangian
(\ref{lagrangian}) we have two possibilities to analyze the mixing and
time evolution of the involved states: the analysis of both
by separating the mixing from the instability or the LOY theory which 
combines both.
Whereas the Lagrangian choice leads to, what we could call standard
properties of the states, the LOY theory does not due to (\ref{nonortho}).
Again we will work here with the neutral kaon system, but keeping in mind
that the very same conclusion would follow had we applied the LOY theory
to (\ref{lagrangian}).

The first curious problem has to do with definition of anti-particles
to $K_{S/L}$. If $\Theta \equiv CPT$ transforms the strangeness eigenstates
as 
$\Theta \vert \bar{K^0} \rangle = e^{-i\delta}\vert K^0 \rangle$ 
\cite{leebook}, then
the CPT-transformed mass-eigenstates are given by
\begin{eqnarray} \label{cptstate}
\vert K_{S/L}^{\Theta} \rangle &=& {e^{i\delta} \over 2pq}
[(\vert p\vert^2 \pm e^{-2i\delta}\vert q\vert^2)\vert K_S \rangle
\nonumber \\
&+& (-\vert p\vert^2 \pm e^{-2i\delta}\vert q\vert^2)\vert K_L \rangle]
\end{eqnarray}
If, up to a phase, we would demand that $\vert K_S^{\Theta}\rangle
=\vert K_S \rangle$ (which is reasonable as the mass-eigenstates kaon
do not carry any other quantum number except mass and spin $0$), we would end 
up with $e^{-2i\delta}=1$ and $\vert p\vert^2 = \vert q\vert^2$ which obviously
is not what one would like to have in the LOY theory. Working with 
non-hermitian operators, there is still a different possibility to define
the CPT-transformation, viz. \cite{rajeev}
\begin{equation} \label{nonhermcpt}
\Theta {\cal M}_{\rm eff} \Theta^{-1} = {\cal M}_{\rm eff}^{\dagger}
\end{equation}
Then one can prove the following theorem \cite{rajeev}: 
provided that the non-hermitian
Hamiltonian ${\cal H}$ is normal i.e $[{\cal H}, {\cal H}^{\dagger}]=0$, there
exist for every state $\vert \Psi \rangle $ with definite mass and lifetime a 
CPT-transformed state defined by 
$\vert \Psi^{\theta} \rangle \equiv \Theta^{-1}\vert \Psi \rangle$ 
with the same
mass and lifetime as $\vert \Psi \rangle$. Certainly, for normal Hamiltonians
this is a good definition of CPT and anti-particle states. Unfortunately in our case we get
\begin{equation} \label{commutator} 
[{\cal M}_{\rm eff}, {\cal M}_{\rm eff}^{\dagger}]= \vert p \vert^2 
-\vert q \vert^2
\end{equation}
which brings us back to the very source of the problem. It seems therefore
that we cannot unambiguously define anti-particle states to the kaon 
mass-eigenstates as long as the latter are non-orthogonal.

The second curious consequence of the non-orthogonality has to do with
time evolution. 
There are at least three different proofs \cite{kha, suder, kabpil}
of the following
statement: 
as long as $\vert p \vert ^2 -\vert q \vert^2 \neq 0$, the
time evolution  beyond the Weisskopf-Wigner approximation is strictly given by
\begin{equation} \label{timekha}
\vert K_{S/L}(\tau)\rangle = p_{SS/LL}(\tau)\vert K_{S/L} \rangle
+ p_{SL/LS}(\tau) \vert K_{L/S} \rangle
\end{equation}
where the coefficients $p_{S/L}=-p_{L/S}$ are non-zero. 
Hence, as long as we insist on 
(\ref{nonortho})
(say,as a signal of CP-violation), we will get a time evolution which looks
like a vacuum regeneration of the mass eigenstates. It has been estimated
that the coefficient $p_{SL}$ is tiny \cite{suder, me2}, 
indeed too small to be detected
experimentally, but as a matter of principle we should be worried about the
interpretation of this effect.

The two last examples display already the unconventional properties of
the non-orthogonality of the mass-eigenstates emerging form the LOY theory.
We point out once again that they would apply also to the dynamics of the 
Lagrangian in section 2, had we applied the LOY theory to the mixing and
instability.

The last example we would like to mention has to do with a paradox and is as 
such less clear. We will discuss it briefly for completeness since 
the paradox would find its simple resolution 
if the mass-eigenstates of kaons were orthogonal. There are actually two types
of paradoxes discussed in the literature in connection with neutral kaons.
Both have to do with entangled kaons at the $\Phi$ resonance. The first one
is a cleverly posed question about future-past interference \cite{yogi}. 
It has been
resolved by means of the unitarity equation (\ref{BS}) and therefore does not 
have anymore the status of paradox \cite{albert}. 
The second one seems to be a genuine
EPR-paradox \cite{datta1}. 
It is, however, based on the wave function collapse of 
non-orthogonal states and on a principle of non-ideal measurements 
\cite{datta2}. Both
issues are not so clear as one would wish them to be so as to be able 
to establish 
unambiguously the existence of the 
paradox. Nevertheless is it worth quoting. The authors of \cite{datta1}
have calculated the probability in an entangled kaon system to measure
a $\bar{K^0}$ on the left hand side at a time $T$ after at $T' < T$
on the right hand side one of the three possibilities has been detected:
(i) strangeness eigenstates $K^0$ or $\bar{K^0}$, (ii) decay products
of $K_S$, (iii) decay products of $K_L$. Their result is
\begin{equation} \label{prob}
{\cal P}(T,T') \propto e^{-\Gamma_S T} + e^{-\Gamma_L T}
-2 \langle K_S \vert K_L \rangle \cos \Delta m (T-T')e^{-{\Gamma_S +
\Gamma_L \over 2}(T+T')}
\end{equation}
Obviously, if correct, 
this is a measureable quantity in which the right hand side
influences the left hand side over arbitrary distances. It therefore violates
the locality principle at a statistical level. This violation has to do
with the $T'$ dependence of (\ref{prob}) which would vanish if
$\langle K_S \vert K_L \rangle =0$.

It is worth mentioning that to avoid certain problems with non-orthogonality
one defines left and right eigenstates of ${\cal M}_{\rm eff}$ \cite{alvarez}
(i.e. we are now dealing  with four states) 
and uses the propagator method for the unstable states \cite{lopez}.
We feel, however, that both methods do not completely answer all the questions
raised by the non-orthogonality. More importantly for the conclusions of the 
present paper, it seems also that both 
these methods also do not agree with the 
Lagrangian one. A more careful examination is required. Such and similar
topic we postpone to future publications. 
    
\setcounter{equation}{0}
\section{Conclusions}
We have emphasized throughout the paper two important features of the
LOY theory. The LOY theory analyzes the mixing and instability 
simultaneously and for the very same reason invokes necessarily a 
non-hermitian effective mass matrix which can lead to non-orthogonality of the
mass-eigenstates. Secondly, this theory is applicable to any two level
system be it elementary or composite. This applies also if the dynamics
is given in form of a Lagrangian. In particular, it is also applicable
to massive neutrinos.
We have presented a special case of a Lagrangian which reveals many
properties of the kaon system, but of course our conclusion is not restricted to this particular case.
In principle, we have then two
non-equivalent choices: to use the LOY theory or to stay within the formalism
of the Lagrangian. The latter defines the mass-eigenstates by unitary
transformations and proceeds to the time evolution of the
individual unstable states in a separate step. We have pointed out some
unusual consequences of the LOY theory.  
In a weaker form we can conclude that if the Lagrangian (\ref{lagrangian})
meets all physical requirements, i.e. we can find a connection between its 
parameters and the parameters of the Standard Model as well as 
the experimentally
measured quantities, we would have equally two valid choices
for the analysis of the neutral kaon system.  

\vskip  2cm

{\elevenbf \noindent Acknowledgments} \newline
The author would like to thank CONACyt for financial support through Catedra
Patrimonial fellowship.

\vskip 2cm

%\end{document}               
                                                         
%\end{references}      

\begin{thebibliography}{99}                                                    
%\begin{references}
\bibitem{ww}
V. F. Weisskopf and E. P. Wigner, Z. Phys.{\bf 63} (1930) 54
\bibitem{loy}
T. D. Lee, R. Oehme and C. N. Yang, Phys. Rev. {\bf 106} (1957) 340
\bibitem{kabir}
P. K. Kabir, {\em The CP Puzzle}, Academic Press, London 1968
\bibitem{petcov}
S. M. Bilenky and S. T. Petcov, Rev. Mod. Phys. {\bf 59} (1987) 671
\bibitem{me1}
For a different scenario of CP-violation with bosons in the Higgs sector see
G. Cvetic, M. Nowakowski and A. Pilaftsis, Phys. Lett. {\bf B301} (1993) 77
\bibitem{bell}
J. S. Bell and J. Steinberger in Proc. Intern. Conf. on Elementary Particles,
Oxford 1965
\bibitem{leebook}
T. D. Lee, {\em Particle Phyisics and Introduction to Field Theory},
Harwood Academic Press 1981
\bibitem{rajeev}
V. S. Mathur and S. G. Rajeev, Mod. Phys. Lett. {\bf A6} (1991) 2741
\bibitem{kha}
L. A. Khalfin, University of Texas at Austin, CPT-Report no. 211 (1990); {\it
ibid} CPT-Report no. 246 (1991)
\bibitem{suder}
C. B. Chiu and E. C. G. Sudarshan, Phys. Rev. {\bf D42} (1990) 3712
\bibitem{kabpil}
P. K. Kabir and A. Pilaftsis, Phys. Rev. {\bf A53} (1996) 66
\bibitem{me2}
M. Nowakowski, Int. J. Mod. Phys. {\bf A14} (1999) 589
\bibitem{yogi} 
Y. Srivastava and A. Widom, Phys. Lett. {\bf B314} (1993) 315
\bibitem{albert}
B. Ancochea and A. Bramon, Phys. Lett. {\bf B347} (1995) 419  
\bibitem{datta1}
A. Datta, D. Home and A. Raychaudhuri, Phys. Lett. {\bf A123} (1987) 4
\bibitem{datta2}
E. Squires and D. Siegwart, Phys. Lett. {\bf A126} (1987) 73;
J. Finkelstein and H. P. Stapp, Phys. Lett. {\bf A126} (1987) 159;
A. Datta, D. Home and A. Raychaudhuri, Phys. Lett. {\bf A130} (1988) 187 
\bibitem{alvarez}
L. Alvarez-Gaume, C. Kounnas, S. Lola and P. Pavlopoulos, 
Phys. Lett. {\bf B458} (1999) 347
\bibitem{lopez}
M. Beuthe, G. Lopez-Castro and J. Pestieau, Int. J. Mod. Phys. {\bf A13} (1998)
3587
\end{thebibliography}
\end{document}